\documentclass[journal]{IEEEtran}
\IEEEoverridecommandlockouts
\usepackage{times,amsmath,color,amssymb,graphicx,epsfig,cite,psfrag,subfigure,algorithm,balance}
\usepackage{amsfonts,pifont,enumerate,cases}
\usepackage{mathrsfs} 
\usepackage[table]{xcolor} 
\usepackage{verbatim}
\usepackage{bm}
\usepackage{cuted,stfloats}
\usepackage{algorithm}
\usepackage{algorithmic}

\usepackage{longtable}
\usepackage{blindtext}
\usepackage{multirow}
\usepackage{float}
\usepackage{threeparttable}
\usepackage{makecell}
\usepackage[utf8]{inputenc}
\usepackage{url}
\usepackage{booktabs}
\usepackage{amssymb}
\usepackage{bbding}
\usepackage{pifont}
\usepackage{wasysym}
\usepackage{utfsym}
\usepackage{fontawesome}
\usepackage[algo2e,ruled,vlined,linesnumbered,lined,boxed,commentsnumbered]{algorithm2e}
\usepackage{amsmath,mathtools}
\usepackage[
    colorlinks=true,
    linkcolor=blue,
    citecolor=blue,
    urlcolor=magenta
]{hyperref}
\usepackage{array}

\begin{document}

\title{Broadband LEO Satellite Constellations for Next-Generation Navigation: Potentials, Enabling Technologies, and Challenges}

\author{Zhendong Li, Mingze Zhu, Jiahao Liu, Zhou Su, Dong Fu, Ruikang Zhong, and Wen Chen
\thanks{Z. Li, M. Zhu, J. Liu and R. Zhong are with the School of Information and Communication Engineering, Xi’an Jiaotong University, Xi’an 710049, China (email: lizhendong@xjtu.edu.cn; mingzezhu@stu.xjtu.edu.cn; xiaochoumeng@stu.xjtu.edu.cn; ruikang.zhong@xjtu.edu.cn). Z. Su is with the School of Cyber Science and Engineering, Xi'an Jiaotong University, Xi'an 710049, China (email: zhousu@ieee.org). D. Fu is with the Northwest Institute of Nuclear Technology, Xi'an 710024, China (email: fudong@nint.ac.cn). W. Chen is with the School of Integrated Circuits, Shanghai Jiao Tong University, Shanghai 200240, China (e-mail: wenchen@sjtu.edu.cn). }
\thanks{(Corresponding author: Zhou Su)}
\vspace{-1.5em}}

\maketitle

\begin{abstract}
The rapid deployment of broadband low Earth orbit (LEO) constellations provides new opportunities for next-generation navigation. Compared with conventional global navigation satellite systems (GNSS), broadband LEO systems provide stronger received signals, wider bandwidths, rapidly varying satellite geometry, and larger constellation scales, offering new capabilities for positioning, navigation, and timing (PNT). Meanwhile, their communication-oriented waveforms, high dynamics, and resource-constrained architectures introduce new challenges that are fundamentally different from those in GNSS. This article provides a comprehensive overview of broadband LEO constellations for next-generation navigation, covering their fundamental characteristics, representative use cases, key enabling technologies, and open challenges. We first compare broadband LEO systems with conventional GNSS and summarize the main technical routes for LEO-based PNT. We then discussed the emerging use cases in GNSS-challenged, multipath-dominant, and integrated communication and navigation (ICAN) scenarios. Key enabling technologies including waveform design, high-dynamic signal processing, and constellation resource scheduling are further reviewed. As a representative  case study, the navigation-oriented optimization of the Starlink primary synchronization sequence demonstrates that subcarrier power allocation can improve multipath resistance while maintaining ranging accuracy. Finally,  several challenges and future directions of broadband LEO-based navigation are discussed, aiming to provide reference for future researches.
\end{abstract}

\section{Introduction}

\IEEEPARstart{P}{ositioning}, navigation, and timing (PNT) services have become indispensable to modern infrastructures, intelligent transportation, and emerging autonomous systems. Global navigation satellite systems (GNSS) remain the primary source of global PNT information. However, GNSS satellites are primarily deployed in medium Earth orbit (MEO), where the long propagation distance results in relatively weak received signals at the user terminal, reducing their robustness to blockage, interference, multipath, and spoofing. These limitations become particularly evident in urban canyons, indoor environments, and other GNSS-challenged scenarios. Meanwhile, the rapid deployment of broadband low Earth orbit (LEO) constellations for global communications has created new opportunities for satellite-based navigation. With lower orbital altitudes, wider bandwidth, larger constellation scales, and faster geometric changes, broadband LEO constellations offer enhanced navigation capabilities that can complement conventional GNSS  \cite{survey_opportunistic_leo_pnt}.

As a typical characteristic of LEO constellations, the lower orbital altitude of LEO satellites provides a more favorable link budget for navigation. In \cite{leo_based_positioning_6g}, the authors highlighted that the shorter propagation distance can provide stronger received signals than conventional GNSS, which is beneficial in environments affected by blockage, attenuation, and interference. Stronger received signals can also improve the reliability of signal acquisition and tracking, especially when propagation losses or local interference degrade conventional navigation signals. This advantage makes existing LEO communication signals attractive for opportunistic navigation. In particular, the authors in \cite{starlink_first_results} developed acquisition and Doppler-tracking methods for Starlink signals, and demonstrated the feasibility of positioning using multiple Starlink satellites.

Wide bandwidth is another important characteristic of broadband LEO systems. A larger signal bandwidth can provide finer delay resolution and improve the potential ranging performance, while offering greater flexibility for navigation-oriented signal design. In \cite{css_leo_pnt}, the authors analyzed a multi-slope chirp spread spectrum waveform for LEO PNT and showed that increasing the signal bandwidth can improve ranging accuracy, although at the cost of higher receiver processing complexity. Beyond dedicated navigation waveform design, broadband LEO communication systems can also support both communication and localization. The authors in \cite{leo_ical_mimo} developed a broadband massive MIMO LEO system for integrated communications and localization, with hybrid precoding designed to balance communication performance and localization accuracy.

Large LEO constellations and fast geometric changes can provide more favorable and rapidly varying satellite geometry for navigation. The authors in \cite{broadband_leo_navigation} analyzed the geometry of broadband LEO constellations and showed that their large number of satellites can substantially improve dilution of precision (DOP) performance compared with conventional navigation constellations. This demonstrates the geometric benefit of large scale LEO deployment. In \cite{multiconstellation_leo_sop}, the authors experimentally combined signals from Starlink, OneWeb, Orbcomm, and Iridium for stationary positioning and mobile navigation, demonstrating the benefit of using satellites distributed across different constellations, frequencies, and orbital configurations. Despite these promising developments, several fundamental and practical aspects of broadband LEO navigation still require systematic investigation.

 Motivated by this, this article provides an overview of broadband LEO constellations for next-generation navigation. Specifically, the fundamentals and major technical routes of broadband LEO navigation are introduced in Section \ref{s2}, while representative use cases are discussed in Section \ref{s3}. In Section \ref{s4}, three key enabling technologies, including signal and waveform design, signal processing, and constellation design and resource scheduling are presented. Section \ref{s5} provides representative case study on navigation-oriented optimization of Starlink signals to demonstrate the performance benefits of signal optimization. The open challenges and future directions are discussed in Section \ref{s6}, and the article is concluded in Section \ref{s7}.

\section{Fundamentals of Broadband LEO Navigation}\label{s2}

\begin{table*}[t]
\centering
\caption{Comparison of GNSS and Representative LEO Communication Constellations}
\label{tab:gnss_leo_comparison}
\renewcommand{\arraystretch}{1.15}
\begin{tabular}{p{2.0cm} p{1.2cm} p{1.8cm} p{2.5cm} p{2.2cm} p{3.2cm}}
\hline
\textbf{System} &
\textbf{Orbit} &
\textbf{Altitude} &
\textbf{Representative Bandwidth} &
\textbf{In-Orbit Satellites} &
\textbf{Primary Mission} \\
\hline

GPS &
MEO &
$\sim$20,200 km &
$\sim$2 MHz (L1 C/A) &
31 operational &
PNT \\

Starlink &
LEO &
$\sim$550 km &
$\sim$240 MHz &
$\sim$11,100 &
Broadband communication \\

OneWeb &
LEO &
$\sim$1,200 km &
$\sim$230 MHz &
$\sim$650 &
Broadband communication \\

Orbcomm &
LEO &
$\sim$700--750 km &
$\sim$4.8 kHz &
58 in orbit (29 working) &
Messaging and IoT communication \\

Iridium &
LEO &
$\sim$780 km &
$\sim$31.5 kHz &
80 (66 operational + 14 spares) &
Voice and data communication \\
\hline
\end{tabular}
\end{table*}

Broadband LEO navigation differs from conventional GNSS not only in orbital altitude, but also in system objective, signal characteristics, and constellation architecture. Before discussing specific applications and key technologies, it is necessary to clarify these differences. This section first compares GNSS with representative LEO communication constellations, then summarizes the main advantages of broadband LEO for navigation, and finally introduces several technical routes of broadband LEO navigation.

\subsection{Fundamental Differences}

GNSS is designed specially for PNT. Its satellites transmit dedicated navigation signals together with orbit, clock, and timing information required by the receiver. In contrast, most LEO constellations are primarily designed for connectivity rather than navigation. Starlink and OneWeb target broadband access, whereas Iridium and Orbcomm support narrower band voice, data, and messaging applications. Consequently, their waveforms, satellite resources, and network operation are optimized primarily for communications rather than PNT.

The different orbital altitude leads to a difference in the radio link. GNSS constellations mainly operate in MEO, while LEO satellites orbit much closer to the Earth. The shorter propagation distance can reduce path loss and provide a received-signal advantage compared with MEO navigation satellites. However, LEO satellites have much smaller coverage footprints. Consequently, more satellites are required to provide stable global coverage. This requirement is compatible with modern communication mega constellations, whose deployment scales are much larger than those of conventional navigation constellations.

The signals themselves are also different. Traditional GNSS relies on dedicated navigation waveforms, while modern broadband LEO systems such as Starlink and OneWeb employ wideband communication signals. In contrast, other LEO communication constellations, such as Iridium and Orbcomm, operate with much narrower signals but can still provide useful measurements for opportunistic navigation. Moreover, communication constellations may use directional beams and dynamically allocate time, frequency, and spatial resources according to communication demand. Consequently, LEO systems generally provide less standardized signal, timing, and satellite state information than conventional GNSS. Although this article focuses on broadband LEO constellations, traditional LEO communication systems such as Iridium and Orbcomm are also included because of their relevance to opportunistic and multisource navigation. Table~\ref{tab:gnss_leo_comparison} summarizes these differences, with Global Positioning System (GPS) included as a representative GNSS for comparison.

\subsection{Main Advantages}

Beyond the favorable link budget associated with lower orbital altitude, the following characteristics are particularly important for broadband LEO navigation.

\textit{Wide Bandwidths:}
A distinctive feature of modern broadband LEO constellations is their large signal bandwidth. Wider bandwidth provides finer delay discrimination, which can improve ranging performance and help distinguish propagation paths with different delays. This is particularly useful in strong multipath environments, where closely spaced signal components may otherwise be difficult to separate. In addition, the combination of high received signal strength and large bandwidth can reduce receiver-noise and multipath contributions to ranging error \cite{fused_leo_gnss}. However, the actual ranging gain also depends on the waveform, spectral shape, signal-to-noise ratio, and receiver processing, rather than bandwidth alone.

\textit{Fast Geometric Changes:}
 Compared with MEO satellites, LEO satellites move rapidly with respect to users on the ground, resulting in large Doppler shifts, noticeable Doppler rates, and rapidly changing satellite-user geometry. This motion produces greater variation in measurement geometry over short observation intervals. Such rapid geometric variation can shorten the carrier-phase ambiguity resolution time in high-precision positioning applications. It can also improve the observability of user states by providing rapidly changing relative-motion information.

\textit{Large Constellations:}
 Large LEO constellations increase the number of satellites passing through a user's field of view and more visible satellites can improve spatial geometry and provide measurement redundancy. \cite{broadband_leo_navigation} analyzed the geometry of broadband LEO constellations and showed that their large satellite populations can achieve favorable dilution-of-precision performance compared with conventional navigation constellations. However, the benefit depends on constellation geometry rather than satellite number alone. Communication constellations are not optimized specifically for navigation, and their DOP may vary with altitude, orbital configuration, and satellite visibility. Thus, large constellation size provides the opportunity for better geometry and redundancy, but the actual gain depends on how the available satellites are selected and used.

\subsection{Technical Routes}

\begin{figure*}[t]
    \centering
    \includegraphics[
        width=0.85\textwidth,
        height=0.38\textheight
    ]{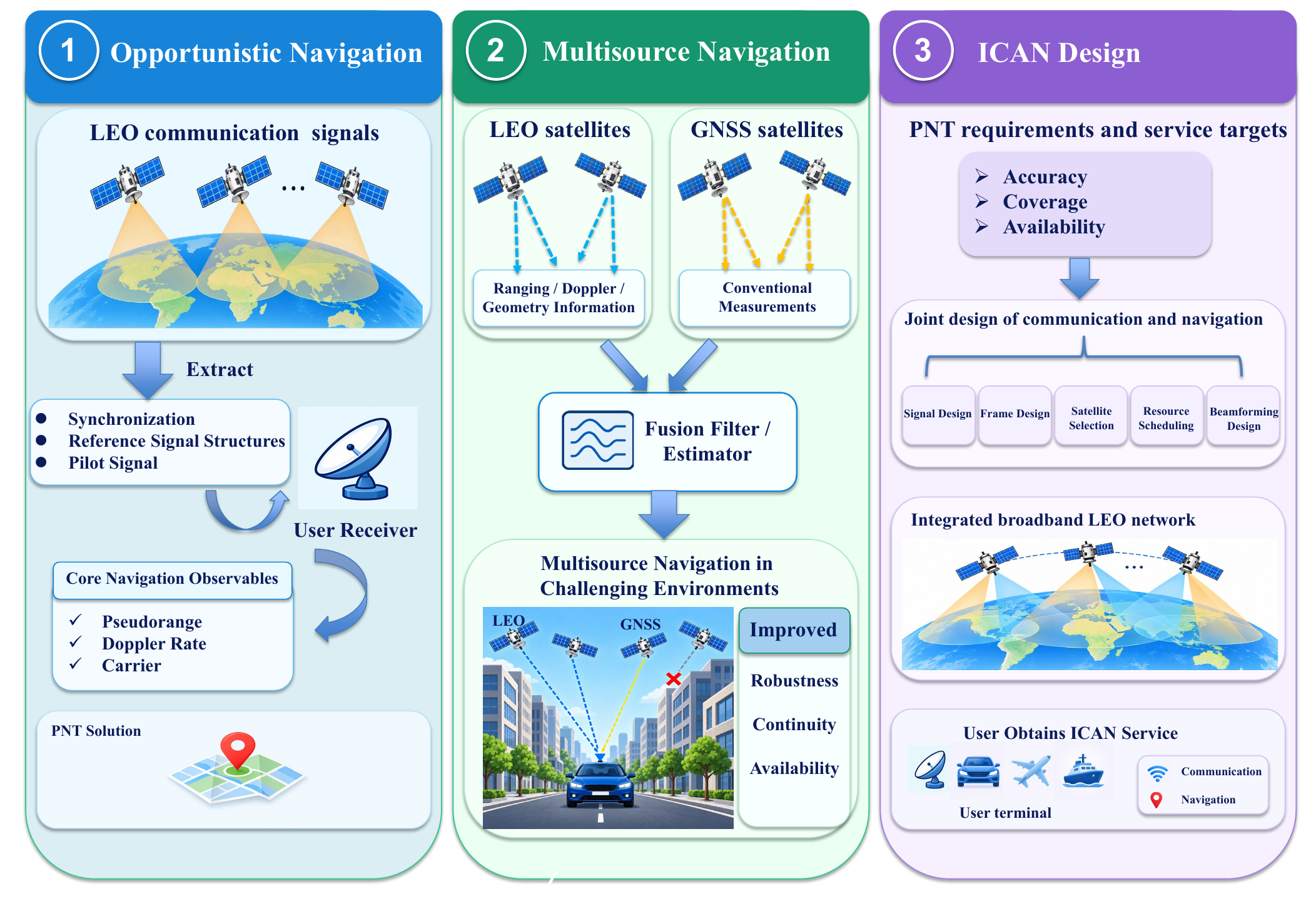}
    \caption{Technical routes for broadband LEO navigation.}
    \label{fig:technical_routes}
\end{figure*}

Current research on broadband LEO PNT mainly follows three technical routes: opportunistic navigation, multisource navigation, and ICAN design. As illustrated in Fig.~\ref{fig:technical_routes}, opportunistic navigation extracts navigation observables from existing communication downlinks at the receiver side, multisource navigation combines complementary measurements from different PNT sources, while ICAN incorporates navigation requirements directly into the design and operation of the communication network. 

\textit{Opportunistic Navigation:}
This route directly exploits existing LEO communication signals for PNT without requiring modifications to the satellite system. Navigation-related parameters are obtained from communication downlinks by making use of their synchronization, reference, or other known signal structures. In \cite{leo_5g_sync}, the authors investigated timing and frequency synchronization for a 5G-integrated LEO satellite communication system and developed estimation and tracking methods for large Doppler conditions. Although the work focuses on communication synchronization, it illustrates that timing and frequency parameters can be continuously estimated from broadband LEO communication signals. Opportunistic navigation follows a similar receiver-side philosophy, with its performance depending strongly on signal knowledge and the availability of satellite orbit and clock information.

\textit{Multisource Navigation:}
LEO-based navigation can also be combined with other navigation sources rather than being used independently. Typical configurations include multi-LEO positioning and the integration of LEO measurements with GNSS, allowing different sources to complement each other in terms of availability, geometry, and measurement diversity. Such fusion is particularly attractive in GNSS-challenged environments, where LEO measurements can provide additional positioning information when GNSS visibility or signal quality is degraded. Multi-LEO positioning and joint LEO-GNSS positioning have been studied in the context of 6G NTN positioning. The results showed that LEO satellites can provide useful positioning support when the number of visible GNSS satellites is limited or GNSS signals are weak \cite{ntn_6g_localization}. This route therefore provides a flexible way to enhance navigation continuity and robustness by combining complementary satellite-based PNT sources.

\textit{ICAN Design:}
A more cooperative route is to incorporate PNT requirements into broadband LEO communication systems from the design stage. Rather than passively extracting navigation information from existing downlinks, communication and navigation are jointly considered when designing signal structures, satellite resources, beamforming, and system operation. The resulting resource coupling creates tradeoffs between communication and navigation performance, requiring coordinated system design. \cite{leo_ican_network} developed an ICAN-enabled LEO framework covering system architecture, key technologies, and unified performance characterization, and jointly optimized satellite selection and beamforming to balance communication and navigation performance. Such designs enable navigation to become an intrinsic capability of broadband LEO communication networks.

\section{Representative Use Cases of Broadband LEO Navigation}\label{s3}

This section discusses representative use cases in which broadband LEO constellations can enhance PNT performance or support ICAN services. As illustrated in Fig. \ref{fig:usecase}, we focus on resilient PNT in GNSS-challenged environments, navigation in multipath-dominant environments, and joint connectivity and PNT services.

\subsection{Resilient PNT in GNSS-Challenged Environments}

Urban canyons, indoor environments, and locally interfered areas can severely degrade GNSS signal availability and measurement continuity. Under weak-signal conditions, reliable signal acquisition and tracking become more challenging, especially when longer integral times and increased computational complexity are required to improve sensitivity.

\begin{figure*}[t]
    \centering
    \includegraphics[
        width=0.90\textwidth,
        height=0.38\textheight
    ]{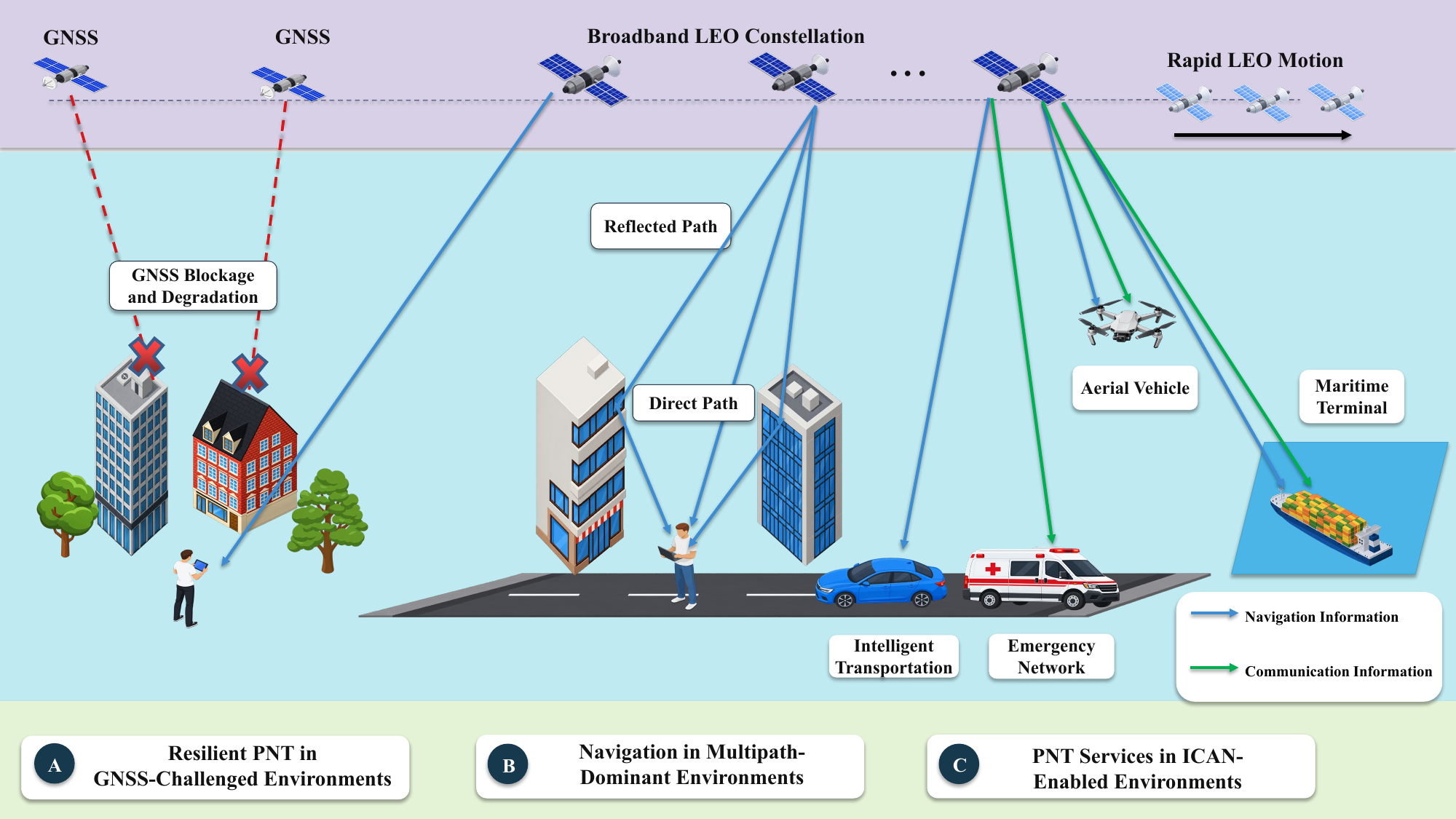}
    \caption{Illustration of representative use cases for broadband LEO navigation.}
    \label{fig:usecase}
\end{figure*}

As illustrated in Fig.~\ref{fig:usecase}, broadband LEO systems can provide a complementary source of PNT information in GNSS-challenged environments. Their lower orbital altitudes result in stronger received signals, while large constellations increase satellite visibility and provide richer satellite--user geometry. These characteristics can improve the availability of navigation signals in environments where GNSS navigation performance is degraded by blockage, attenuation, or interference. In such scenarios, broadband LEO is better viewed as a complementary PNT layer rather than a replacement for GNSS, providing additional measurements that can improve service availability, continuity, and resilience.

\subsection{Navigation in Multipath-Dominant Environments}

In dense urban areas, building corridors, and indoor environments, reflections and non-line-of-sight (NLoS) propagation introduce delayed signal components that distort correlation peaks and degrade time-of-arrival and ranging measurements. The relatively large bandwidth of broadband LEO signals provides finer delay resolution, offering greater potential to distinguish propagation paths with different delays \cite{broadband_leo_navigation}. In addition, the rapid motion of LEO satellites causes the satellite-user-reflector geometry to vary more quickly than in conventional MEO-based systems, which can reduce the persistence of multipath effects over short observation intervals. These characteristics provide opportunities to improve the robustness of LEO-based navigation in multipath-dominant environments, particularly when combined with appropriate signal processing and waveform design techniques.

\subsection{PNT Services in ICAN-Enabled Environments}

Broadband LEO constellations are primarily deployed to provide high-capacity connectivity, while many emerging users, including aerial vehicles, intelligent transportation systems, maritime terminals, and emergency networks, also require continuous PNT services. Supporting connectivity and PNT through the same satellite infrastructure can improve resource utilization and reduce the need for separate service platforms. However, communication and navigation may share satellite, beam, power, and other radio resources, resulting in inherent performance coupling. In \cite{leo_ican_twc}, satellite selection and beamforming were jointly optimized for a LEO-ICAN network, with transmission rate and geometric dilution of precision (GDOP) used to characterize communication and navigation performance, respectively. The results demonstrated the tradeoff between the two services and highlighted the need for coordinated resource allocation. Through such coordination, broadband LEO networks can jointly support connectivity and PNT services, making them suitable for scenarios with integrated communication and navigation requirements.

%  后向兼容性

\section{Key Enabling Technologies for Broadband LEO Navigation}\label{s4}

The navigation capability of broadband LEO constellations depends not only on their favorable link budget and constellation geometry, but also on how communication signals, receiver algorithms, and system resources are adapted for PNT. Unlike conventional GNSS, these systems are primarily designed for communications, making navigation closely coupled with the underlying waveform, high-dynamic signal processing, and constellation resource management. The following technologies represent three important layers for exploiting broadband LEO systems for navigation.

\subsection{Signal and Waveform Design}
Broadband waveform design provides considerable flexibility for navigation-oriented signal optimization. Signal bandwidth, spectral shape, and power distribution can directly affect ranging accuracy, multipath resistance, interference robustness, and spectral compatibility. These design degrees of freedom are particularly relevant to multicarrier broadband LEO signals, whose spectral resources can be flexibly configured for PNT purposes. For instance, the Starlink Ku-band downlink has been identified as a broadband orthogonal frequency division multiplexing (OFDM) signal with predictable synchronization sequences that can support pseudorange and Doppler measurements \cite{starlink_signal_structure}. Its multicarrier structure further enables navigation-oriented optimization by adjusting the power allocation across subcarriers to reshape the signal spectrum and correlation characteristics. Beyond adapting existing communication waveforms, communication and navigation signals can also be jointly designed. Navigation PRN codes and communication data can be assigned to orthogonal resources in the delay-Doppler domain, reducing mutual interference and allowing navigation receivers to process only the navigation-related bandwidth. Such waveform designs can support continuous navigation tracking while maintaining reliable communication in highly dynamic LEO channels.

\subsection{Signal Processing}
The high velocity of LEO satellites results in large and rapidly varying Doppler shifts, which enlarge the acquisition search space and increase the complexity of signal synchronization and tracking. Advanced receiver processing is therefore required to maintain reliable operation under such highly dynamic conditions. A Zak-transform-based coherent acquisition method exploits the delay-Doppler structure of periodic navigation signals to mitigate coherent integration loss caused by Doppler mismatch and improve acquisition efficiency \cite{zak_leo_acquisition}. More generally, delay-Doppler domain processing provides an effective means of handling large frequency offsets in LEO navigation. Efficient tracking algorithms are also important for reducing receiver complexity and maintaining measurement continuity during rapid satellite motion. Particularly, the tracking loop must adapt to the fast variation of carrier frequency, code phase, and Doppler rate to avoid loss of lock and preserve stable navigation measurements. In addition, data-driven methods can assist signal detection, measurement-quality assessment, and NLoS identification when accurate signal or propagation models are difficult to obtain. These methods can complement conventional model-based processing and improve receiver robustness in complex LEO environments.

\subsection{Constellation Design and Resource Scheduling}
Large LEO constellations can provide favorable geometry and measurement redundancy, but navigation performance depends on which satellites, beams, and communication resources are actually available to a user. Time, frequency, transmit power, and beam resources may have to be shared between communication and navigation, creating a tradeoff between communication throughput and PNT performance. Navigation-oriented resource scheduling can therefore select satellites and allocate resources according to visibility, geometry, signal quality, and service demand rather than simply maximizing the number of active links. Such scheduling also needs to adapt to the rapid changes in satellite visibility and user-satellite geometry. For example, a practical approach is to dynamically allocate available satellite resources among different navigation tasks according to time-varying visibility and system constraints. Such system-level scheduling connects signal and receiver technologies with the continuity and efficiency required for practical broadband LEO PNT services.

\section{Case Study: Navigation-Oriented Optimization of Starlink Signals}\label{s5}

The original Starlink signal provides favorable ranging potential because of its wide bandwidth, but its communication-oriented spectrum is not ideal for navigation. Taking the Starlink primary synchronization sequence (PSS) as an example, it has an approximately flat in-band spectrum with sharp transitions near the band edges. This spectral shape produces a narrow autocorrelation main peak, but it also introduces noticeable sidelobe oscillations and a relatively long tail. As a result, the signal performs well for short-delay multipath, while larger errors can appear at medium and long delays. This behavior motivates navigation-oriented spectrum shaping without changing the basic OFDM structure.

Following the reported Starlink signal structure \cite{starlink_signal_structure}, the signal bandwidth is set to 240~MHz. For comparison, two representative GNSS modulation schemes are considered. Binary phase-shift keying (BPSK) is widely used in conventional GNSS signals, such as the GPS L1 C/A signal, while binary offset carrier (BOC) modulation is adopted in several modernized GNSS signals, including those of GPS and Galileo. To provide a fair comparison with the 240~MHz Starlink signal, scaled BPSK(120) and BOC(60,60) waveforms are considered. BPSK(120) denotes a BPSK waveform with a chip rate of 120~Mcps, corresponding to a main-lobe bandwidth of approximately 240~MHz. For BOC(60,60), the two parameters denote the subcarrier frequency and code rate, respectively, both set to 60~MHz (with the code rate expressed in Mcps). These parameter settings are used to construct bandwidth-comparable reference waveforms rather than to represent specific operational GNSS signals. The receiver front-end bandwidth is set to 250~MHz. The coherent integration time and loop noise bandwidth are 1~ms and 0.2~Hz, respectively, and the early-late correlator spacing is set to 0.1~chip. For the multipath evaluation, the amplitude ratio between the multipath and direct signals is set to $-6$~dB.

\begin{figure}[H]
    \centering
    \includegraphics[width=0.90\linewidth]{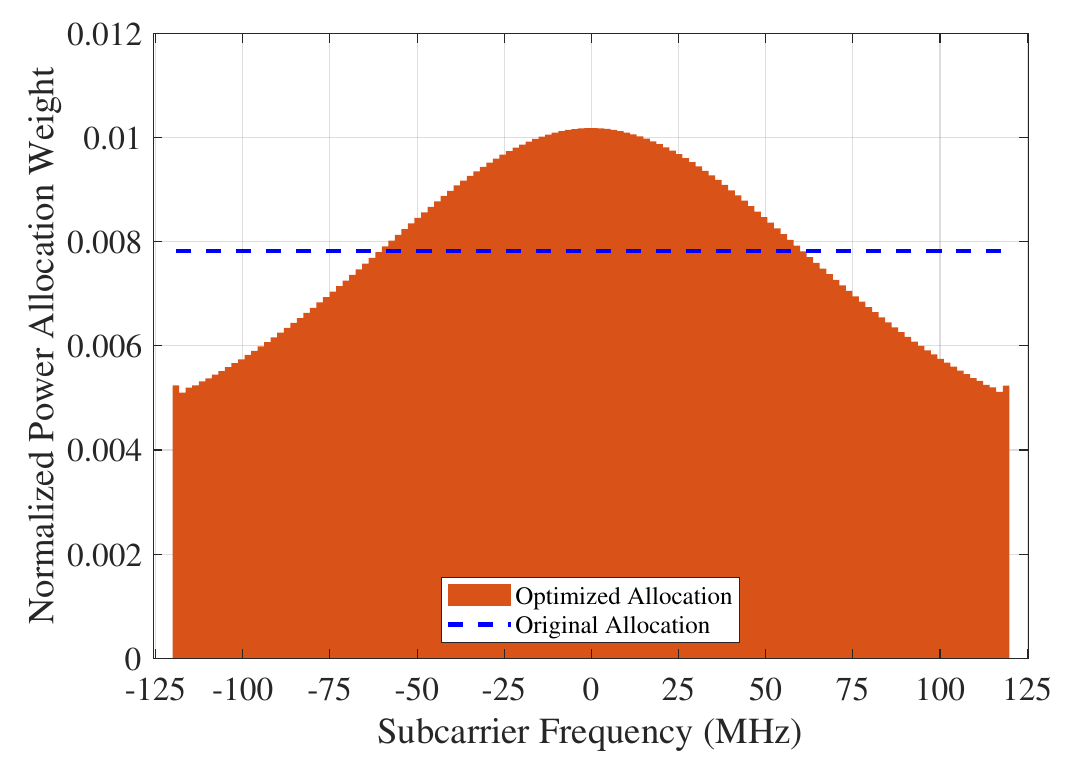}
    \caption{Comparison of power allocation before and after optimization.}
    \label{fig:3}
    \vspace{1.0em}
\end{figure}

Using the Starlink PSS as an example, the subcarrier power allocation is optimized to improve its navigation performance. As shown in Fig.~\ref{fig:3}, the original PSS adopts an approximately uniform power allocation across the occupied subcarriers, whereas the optimized allocation exhibits a smoother, center-weighted profile. Specifically, less power is assigned to the subcarriers near the band edges, while more power is concentrated around the center of the occupied band. This redistribution softens the sharp spectral transitions of the original signal and consequently suppresses the sidelobes of its autocorrelation function. Meanwhile, sufficient Gabor bandwidth is retained to preserve the ranging capability of the wideband signal. The resulting power profile reflects a tradeoff between multipath resistance and ranging accuracy rather than simply maximizing one individual performance metric.

\begin{figure}[H]
    \centering

    \begin{minipage}[t]{0.48\linewidth}
        \centering
        \includegraphics[width=\linewidth]{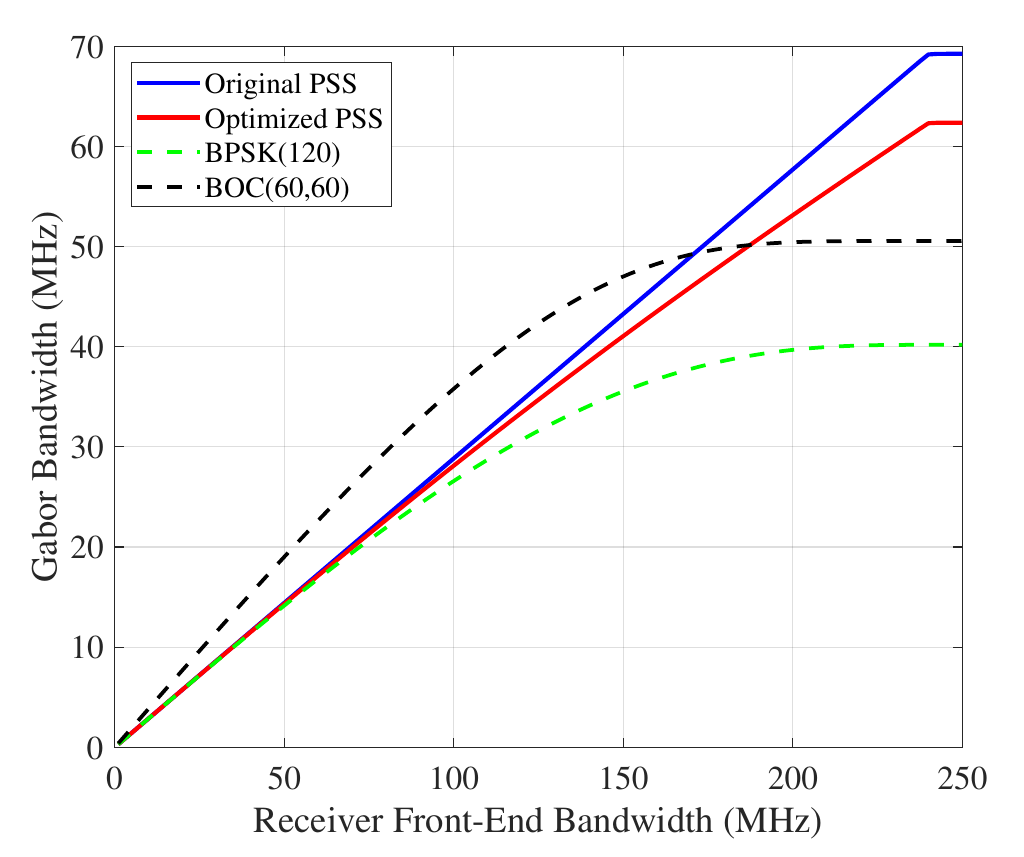}

        \vspace{0.1mm}
        {\footnotesize (a) Gabor bandwidths.}
    \end{minipage}
    \hfill
    \begin{minipage}[t]{0.50\linewidth}
        \centering
        \includegraphics[width=\linewidth]{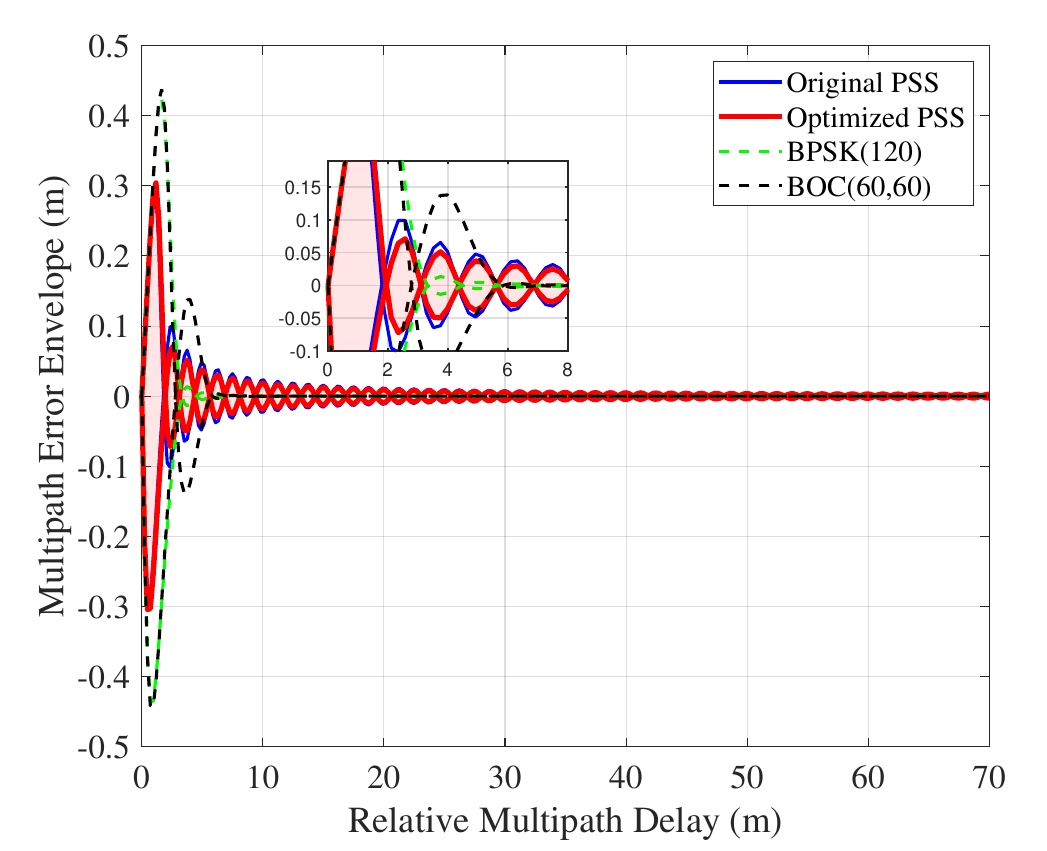}

        \vspace{0.1mm}
        {\footnotesize (b) Multipath error envelopes.}
    \end{minipage}

    \caption{Performance comparison of the optimized signal.}
    \label{fig:4}
\end{figure}

Fig.~\ref{fig:4} further evaluates the optimized waveform from the perspectives of ranging accuracy and multipath resistance. Fig.~\ref{fig:4}(a) compares the Gabor bandwidths of the considered signals as the receiver front-end bandwidth increases. The Gabor bandwidth characterizes the effective spectral spread of a signal, and a larger value generally corresponds to a lower Cramér-Rao lower bound (CRLB) for delay estimation and thus better ranging accuracy. As expected, reducing the power near the band edges slightly decreases the Gabor bandwidth of the optimized Starlink PSS compared with the original signal. Nevertheless, at the 250~MHz receiver front-end bandwidth adopted in the simulation, the optimized PSS still maintains a larger Gabor bandwidth than the considered GNSS reference waveforms, indicating that its ranging advantage is largely preserved. Fig.~\ref{fig:4}(b) compares the corresponding multipath error envelopes. The optimized PSS exhibits substantially weaker oscillations than the original waveform, particularly at medium and long relative delays. Overall, the optimization successfully improves multipath resistance while maintaining strong ranging performance.

\section{Open Challenges and Future Directions}\label{s6}

\subsection{Robust Receiver Processing}

LEO satellites operate at high velocities, resulting in large Doppler shifts and rapid Doppler variations that place increased demands on signal acquisition and tracking. The large Doppler range expands the search space during signal acquisition, while rapidly varying carrier frequency, phase, and code timing make stable tracking more challenging. Meanwhile, the wide bandwidth of broadband LEO signals imposes a substantial processing burden on the receiver, particularly when signals from multiple satellites are processed simultaneously. Existing approaches can improve receiver adaptability to LEO dynamics through Doppler compensation, auxiliary information, and advanced tracking loops, but these techniques may introduce additional computational overhead and implementation complexity. In practical systems, it remains challenging to maintain robust acquisition and tracking while satisfying the real-time processing and complexity constraints of Broadband LEO PNT receivers. Future research should therefore explore the use of dynamic aiding information to reduce acquisition search overhead, together with efficient tracking techniques and low-complexity receiver architectures, to improve real-time operation under highly dynamic conditions.

\subsection{Communication-Navigation Resource Tradeoffs}

In ICAN systems based on broadband LEO constellations, communication and navigation services need to share limited system resources while satisfying their respective performance requirements, resulting in a highly coupled resource allocation problem. Multi-satellite asynchronous transmission further complicates resource coordination and interference management. Existing work has investigated multi-satellite asynchronous transmission by jointly allocating subcarriers and power to coordinate communication and navigation performance while mitigating mutual communication-navigation interference and inter-satellite interference associated with asynchronous transmission \cite{wang_resource_2025}. However, such resource allocation methods are typically designed for specific system configurations and optimization objectives, and their adaptability to dynamic link conditions and diverse communication and navigation service requirements remains limited. In practical broadband LEO systems, it remains challenging to adaptively coordinate communication and navigation resources according to dynamic changes in link conditions and service requirements while maintaining an appropriate balance among different service demands. Future work can investigate joint resource allocation schemes for dynamic multi-satellite scenarios, with real-time link conditions and service requirements incorporated into the resource adjustment process.

\subsection{Trustworthy PNT Services}

Interference, spoofing, insufficient signal authentication, and differences in the quality and error characteristics of heterogeneous observations mean that high positioning accuracy alone does not necessarily guarantee trustworthy PNT services. Interference and spoofing may introduce anomalies into navigation information, while the lack of effective signal authentication makes it more difficult to verify the reliability of navigation information sources. Meanwhile, variations in observation quality and error characteristics across satellites and information sources further complicate anomaly detection and consistency validation across multiple information sources. Therefore, broadband LEO systems need to address not only positioning accuracy, but also service integrity, authentication, and continuity, together with mechanisms for anomaly detection and cross-source consistency validation. Under complex and dynamic conditions, effectively integrating these capabilities while continuously providing trustworthy and verifiable PNT information in the presence of changing observation quality and potential interference remains an open issue. Further research is needed to develop trustworthy PNT processing mechanisms for heterogeneous multi-source observations, combining integrity monitoring, authentication, anomaly detection, and consistency validation to improve the trustworthiness and continuity of PNT services.

\section{Conclusion}\label{s7}

In this article, we provided an overview of broadband LEO satellite constellations for next-generation navigation. We first discussed the fundamental differences between broadband LEO systems and conventional GNSS. Then we introduced the main technical routes of broadband LEO navigation, followed by representative use cases and key enabling technologies. The Starlink PSS case study further demonstrated how broadband communication signals can be adapted to enhance navigation performance. By optimizing the subcarrier power allocation, the spectral energy distribution was reshaped to achieve a more favorable tradeoff between ranging capability and robustness. These results indicate that waveform optimization can improve the ranging robustness and navigation usability of broadband LEO signals without fundamentally changing the underlying signal structure. Finally, several open challenges and future research directions were discussed, including robust receiver processing, communication-navigation resource coordination, and trustworthy PNT services. Further advances in these areas can enhance the robustness, availability, accuracy, and trustworthiness of broadband LEO-enabled PNT, supporting the development of next-generation navigation.

\balance

\bibliographystyle{IEEEtran}
\bibliography{reference}
\vspace{2em}

\end{document}